\documentclass[prl,twocolumn,10pt]{revtex4}

\usepackage{graphicx}
\newcommand{\Version}       {Version: stopping.tex, March 3, 2004}
\newcommand{\bnl}           {$\rm^{1}$}
\newcommand{\ires}          {$\rm^{2}$}
\newcommand{\kraknuc}       {$\rm^{3}$}
\newcommand{\krakow}        {$\rm^{4}$}
\newcommand{\baltimore}     {$\rm^{5}$}
\newcommand{\newyork}       {$\rm^{6}$}
\newcommand{\nbi}           {$\rm^{7}$}
\newcommand{\texas}         {$\rm^{8}$}
\newcommand{\bergen}        {$\rm^{9}$}
\newcommand{\bucharest}     {$\rm^{10}$}
\newcommand{\kansas}        {$\rm^{11}$}
\newcommand{\oslo}          {$\rm^{12}$}

\def\drap{\langle \delta y \rangle}
\def\dbar{\langle y \rangle}

\def\mpt{\langle p_T \rangle}
\def\mmt{\langle m_T \rangle}

\def\Nb{N_{B-\bar{B}}}

\def\nn{n_n}
\def\np{n_p}
\def\nl{n_\Lambda}
\def\nb{n_B}
\def\npm{n_{p,meas}}
\def\nsp{n_{\Sigma^+}}
\def\nsm{n_{\Sigma^-}}

\def\nnp{\nn/\np}
\def\nlp{\nl/\np}
\def\nspp{\nsp/\np}
\def\nsmp{\nsm/\np}

\newcommand{\rootsnn}[1]{$\sqrt{s_{NN}} = #1$ GeV}

\begin{document}

\title{Nuclear Stopping in Au+Au Collisions at \rootsnn{200}}

\date{\Version}

\author{
  I.~G.~Bearden\nbi, 
  D.~Beavis\bnl, 
  C.~Besliu\bucharest, 
  B.~Budick\newyork, 
  H.~B{\o}ggild\nbi, 
  C.~Chasman\bnl, 
  C.~H.~Christensen\nbi, 
  P.~Christiansen\nbi, 
  J.~Cibor\kraknuc, 
  R.~Debbe\bnl, 
  E. Enger\oslo,  
  J.~J.~Gaardh{\o}je\nbi, 
  M.~Germinario\nbi, 
  K.~Hagel\texas, 
  O.~Hansen\nbi, 
  A.~Holm\nbi, 
  A.~K.~Holme\oslo, 
  H.~Ito\kansas$^{,}$\bnl, 
  A.~Jipa\bucharest, 
  F.~Jundt\ires, 
  J.~I.~J{\o}rdre\bergen, 
  C.~E.~J{\o}rgensen\nbi, 
  R.~Karabowicz\krakow, 
  E.~J.~Kim\bnl$^{,}$\kansas, 
  T.~Kozik\krakow, 
  T.~M.~Larsen\oslo, 
  J.~H.~Lee\bnl, 
  Y.~K.~Lee\baltimore, 
  G.~L{\o}vh{\o}iden\oslo, 
  Z.~Majka\krakow, 
  A.~Makeev\texas, 
  M.~Mikelsen\oslo, 
  M.~Murray\texas$^{,}$\kansas, 
  J.~Natowitz\texas, 
  B.~S.~Nielsen\nbi, 
  J.~Norris\kansas, 
  K.~Olchanski\bnl, 
  D.~Ouerdane\nbi, 
  R.~P\l aneta\krakow, 
  F.~Rami\ires, 
  C.~Ristea\bucharest, 
  D.~R{\"o}hrich\bergen, 
  B.~H.~Samset\oslo, 
  D.~Sandberg\nbi, 
  S.~J.~Sanders\kansas, 
  R.~A.~Scheetz\bnl, 
  P.~Staszel\nbi$^{,}$\krakow, 
  T.~S.~Tveter\oslo, 
  F.~Videb{\ae}k\bnl, 
  R.~Wada\texas, 
  Z.~Yin\bergen, and 
  I.~S.~Zgura\bucharest\\ 
  The BRAHMS Collaboration \\ [1ex]
  \bnl~Brookhaven National Laboratory, Upton, New York 11973\\
  \ires~Institut de Recherches Subatomiques and Universit{\'e} Louis
  Pasteur, Strasbourg, France\\
  \kraknuc~Institute of Nuclear Physics, Krakow, Poland\\
  \krakow~Smoluchkowski Inst. of Physics, Jagiellonian University,
  Krakow, Poland\\
  \baltimore~Johns Hopkins University, Baltimore 21218\\
  \newyork~New York University, New York 10003\\
  \nbi~Niels Bohr Institute, Blegdamsvej 17, University of Copenhagen,
  Copenhagen 2100, Denmark\\
  \texas~Texas A$\&$M University, College Station, Texas, 17843\\
  \bergen~University of Bergen, Department of Physics, Bergen,
  Norway\\
  \bucharest~University of Bucharest, Romania\\
  \kansas~University of Kansas, Lawerence, Kansas 66045\\
  \oslo~University of Oslo, Department of Physics, Oslo, Norway\\
}

\begin{abstract}
  Transverse momentum spectra and rapidity densities, $dN/dy$, of
  protons, anti--protons, and net--protons ($p-\bar{p}$) from central
  (0-5\%) Au+Au collisions at \rootsnn{200} were measured with the
  BRAHMS experiment within the rapidity range $0 \leq y \leq 3$. The
  proton and anti--proton $dN/dy$ decrease from mid--rapidity to
  $y=3$.  The net--proton yield is roughly constant for $y<1$ at
  $dN/dy \sim 7$, and increases to $dN/dy \sim 12$ at $y \sim 3$. The
  data show that collisions at this energy exhibit a high degree of
  transparency and that the linear scaling of rapidity loss with
  rapidity observed at lower energies is broken.  The energy loss per
  participant nucleon is estimated to be $73 \pm 6$ GeV.
  
\end{abstract}

\maketitle

The energy loss of colliding nuclei is a fundamental quantity
determining the energy available for particle production (excitation)
in heavy ion collisions. This deposited energy is essential for the
possible formation of a deconfined quark--gluon phase of matter (QGP).
Because baryon number is conserved, and rapidity distributions are
only slightly affected by rescattering in late stages of the
collision, the measured net--baryon ($B-\bar{B}$) distribution retains
information about the energy loss and allows the degree of nuclear
stopping to be determined. Such measurements can also distinguish
between different proposed phenomenological mechanisms of initial
coherent multiple interactions and baryon
transport~\cite{Bass:1999zq,Soff:2002bn,Bass:2002vm} .

The average rapidity loss, $\drap = y_p - \dbar$~\footnote{The
rapidity, $y$, is defined as $y=0.5\ln{\frac{E+p_z}{E-p_z}}$. Rapidity
variables are in the center-of-mass system with $y_p$ positive.}, is
used to quantify stopping in heavy ion
collisions~\cite{Busza:1983rj,Videbaek:mf}. Here, $y_p$ is rapidity of
the incoming projectile and $\dbar$ is the mean net--baryon rapidity
after the collision :

\begin{equation}
  \label{eq:stop:baryonafter}
  \dbar = \frac{2}{N_{part}} 
  \int^{y_p}_{0} y \cdot \frac{dN_{(B-\bar{B})}(y)}{dy} \cdot dy,
\end{equation}

\noindent
where $N_{part}$ is the number of participating nucleons in the
collision. The two extremes correspond to full stopping, where initial
baryons lose all kinetic energy ($\drap = y_p$) and full transparency,
where they lose no kinetic energy ($\drap= 0$).  For fixed collision
geometry (system size and centrality) at lower energy (SIS, AGS, and
SPS) it was observed that $\drap$ is proportional to the
projectile rapidity. For central collisions between heavy nuclei (Pb,
Au), $\drap \sim 0.58 \cdot y_p$~\cite{Videbaek:mf,Hong:1997mr,e917}.

Bjorken assumed that sufficiently high energy collisions are
``transparent'', thus the mid--rapidity region is approximately
net--baryon free ~\cite{Bjorken:1982qr}. The energy density early in
the collision, $\epsilon$, can then be related in a simple way to the
final particle production. At RHIC it has been estimated that
$\epsilon \sim 5 \text{ GeV/fm}^3$, well above the lattice QCD
prediction ($\epsilon_{crit} \sim 1
\text{GeV/fm}^3$~\cite{Karsch:2001vs}) for the hadron gas to QGP phase
transition.

In this letter, results on proton and anti--proton production, and
baryon stopping in $Au+Au$ collisions at \rootsnn{200} are presented.
The data were collected with the BRAHMS detector at RHIC. The BRAHMS
experiment consists of two independent spectrometer arms, the
Mid-Rapidity Spectrometer (MRS) and the Forward Spectrometer (FS),
described in detail in~\cite{BRAHMSNIM}. The spectrometers consist of
dipole magnets, Time Projection Chambers (TPC) and Drift Chambers (DC)
for tracking charged particles, and detectors for particle
identification (PID). The MRS can be rotated $30^\circ < \theta <
95^\circ$ and the FS $2.3^\circ < \theta < 30^\circ$, where $\theta$
is the polar angle with respect to the beam axis. By combining
different settings of angle and magnetic fields, (anti--)proton
transverse momentum spectra at different rapidities ($0 \leq y \leq
3$) were obtained.

The interaction point (IP) is determined by the Beam--Beam counters
(BB), with a precision of $\sigma_{BB} = 0.7$ cm. The IP was required
to be within 15 (20) cm of the nominal IP for the MRS (FS) analysis to
minimize acceptance corrections. 

Protons and anti--protons are identified using Time Of Flight
detectors (TOF) in the MRS (TOFW) and FS (H1 and H2), and by the Ring
Imaging Cherenkov (RICH) in the FS. The TOF resolution is
$\sigma_{TOFW} = 80 \text{ ps}$, $\sigma_{H1} = \sigma_{H2} = 90
\text{ ps}$. To identify (anti--)protons it is required that the
derived square of the mass, $m^2$, is within $\pm 2 \sigma$ of the
proton $m^2$ and at least 2$\sigma$ away from the $m^2$ for kaons.
This allows (anti--)protons to be identified up to momentum $p < 3.0$
GeV/$c$ for TOFW, $p < 4.5$ GeV/$c$ for H1, and $p < 6.5$ GeV/$c$ for
H2. The refractive index of the RICH, $n = 1.0020$ allows protons
to be directly identified via a measurement of their ring radius in
the range $15 < p < 25$ GeV/$c$. By using the RICH to veto pions and
kaons the PID can be extended down to $p = 10 \text{ GeV/$c$}$.

The collision centrality is determined using a multiplicity array
located around the nominal IP~\cite{BRAHMSNIM}. In this analysis only
the 0-5\% most central collisions were used. The mean number of
participants for this centrality class was found by Glauber model
calculations to be $N_{part} = 357 \pm 8$~\cite{Bearden:2001qq}.

From the identified particles, invariant differential yields
$\frac{1}{2\pi p_T}\frac{d^2N}{dy dp_T}$ were constructed for each
spectrometer setting. The differential yields were corrected for
geometrical acceptance, tracking and PID efficiency, absorption and
multiple scattering. The corrections for absorption and multiple
scattering were less than 20\% at the lowest $p_T$ and less than 5\%
at the highest $p_T$ in all settings. No corrections were applied for
secondary protons from e.g. the beam pipe, since the contribution was
found to be negligible in GEANT based Monte Carlo (MC)
simulations~\cite{geant} using HIJING~\cite{Wang:1991ht} as input,
when the tracks were required to point back to the IP. The acceptance
correction is purely geometrical and calculated using a MC simulation
of the BRAHMS detector. The tracking efficiency was estimated for the
TPCs, by embedding simulated tracks in real events (method I), and in
the FS where there are 5 tracking detectors (2 TPCs and 3 DCs) by
comparing the number of identified track segments in the chamber to
the number of reference tracks determined by other detectors
disregarding the chamber under consideration (method II). In the front
part of the FS (FFS) where both methods are applicable, there was a
10\% discrepancy between the two methods which has been included in
the systematic errors quoted below.  For the spectrometers the total
tracking efficiency depends on the spectrometer angle and is 90-95\%
in the MRS, 80-90\% in the FFS (settings at $y \sim 2$) and 60-70\% in
the full FS ($y \sim 2$, $y \sim 3$). The efficiencies of the TOF
detectors were found to be 93-98\%. For the part of the spectrum at $y
\sim 3$, where the RICH was used to veto pions and kaons a correction
for contamination of pions and kaons with no identified ring radius
was applied. The RICH inefficiency is $5\%$.

\begin{figure}[htb]
  \includegraphics[width=\columnwidth]{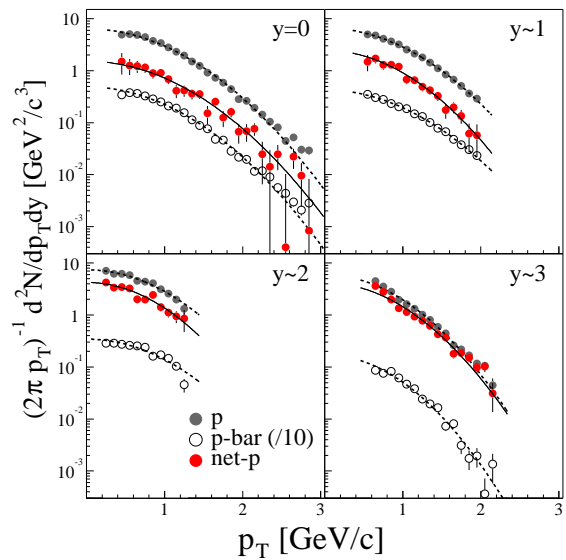}
  \caption{Proton, anti--proton, and net--proton transverse momentum
    spectra at selected rapidities, $y \sim 0, 1, 2, 3$. The solid and
    dashed lines represents Gaussian fits to the data (extended
    outside the fit--range for clarity). The anti--proton spectra have
    been divided by 10 as indicated. The error on the data points are
    statistical only and no weak decay correction has been applied.}
  \label{fig1}
\end{figure}

Figure~\ref{fig1} shows transverse momentum spectra for four of the
nine measured rapidities. The net--proton spectra were constructed by
subtracting the anti--proton spectrum from the proton spectrum. At
each rapidity, the proton, anti--proton, and net--proton spectra have
similar shape, indicating that produced and transported protons have
similar spectral properties. To obtain rapidity densities, $dN/dy$,
for $p, \bar{p}$ and net-$p$ their spectra are fitted, and the fit was
used to extrapolate to the full $p_T$ range.  Different functional
forms were tested: $m_T$-exponential, Boltzmann and Gaussian.  The
function found to best describe the data was the Gaussian in $p_T$
[$f(p_T) \propto e^{-p_T^2/(2\sigma^2)}$] and this function has been
used for all fits. This functional form was also used
in~\cite{Adler:2001aq}.

The mean transverse momentum $\mpt$ of the spectra calculated from the
fit is found to be within 0.1 GeV/$c$ at each rapidity for the three
functional forms. For protons which have the best counting statistics,
$\mpt$ decreases from $\mpt = 1.01 \pm 0.01 \text{(stat) GeV/$c$}$ at
$y=0$ to $\mpt = 0.84 \pm 0.01 \text{(stat) GeV/$c$}$ at $y \sim 3$.

\begin{figure}[htb]
  \includegraphics[width=\columnwidth,clip=]{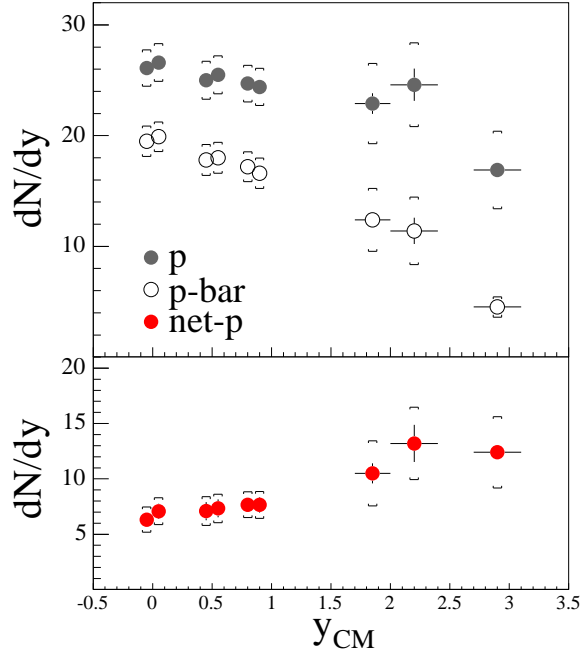}
  \caption{Proton, anti--proton, and net--proton rapidity densities
    $dN/dy$ as a function of rapidity at \rootsnn{200}. The horizontal
    bars shows the rapidity intervals for the projections. The errors
    shown with vertical lines are statistical only while the caps
    includes both statistical and systematic. No weak decay correction
    has been applied.}
 \label{fig2}
\end{figure}

The differential yield within the measured $p_T$ range varies from
85\% of the total $dN/dy$ near mid--rapidity to 45\% at $y \sim 3$.
The systematic errors on $dN/dy$ were estimated from the difference in
$dN/dy$ values obtained using different spectrometer settings covering
the same $(y, p_T)$ regions, the discrepancy between the two different
efficiency methods, and by estimating the effects of the $p_T$
extrapolation. The systematic errors were found to be 10-15\% for
mid-rapidity ($y < 1$) and 20-30\% for forward rapidities.

Figure~\ref{fig2} shows the resulting rapidity densities $dN/dy$ as a
function of rapidity. The most prominent feature of the data is that
while the proton and anti--proton $dN/dy$ decrease at rapidities away
from mid--rapidity the net--proton $dN/dy$ increases over all three
units of rapidity, from $dN/dy\text{($y$=0)} = 6.4 \pm 0.4
\text{(stat)} \pm 1.0 \text{(syst)}$ to $dN/dy\text{($y$=3)} = 12.4
\pm 0.3 \text{(stat)} \pm 3.2 \text{(syst)}$.

A Gaussian fit to the anti-proton $dN/dy$ distribution gives the total
extrapolated anti--proton yield : $84 \pm 6$ (92\% in $-3 < y < 3$).
For protons the yield from a Gaussian fit to $dN/dy$ in the range, $-3
< y < 3$, is $138 \pm 7$.

\begin{figure}[htb]
  \includegraphics[width=\columnwidth]{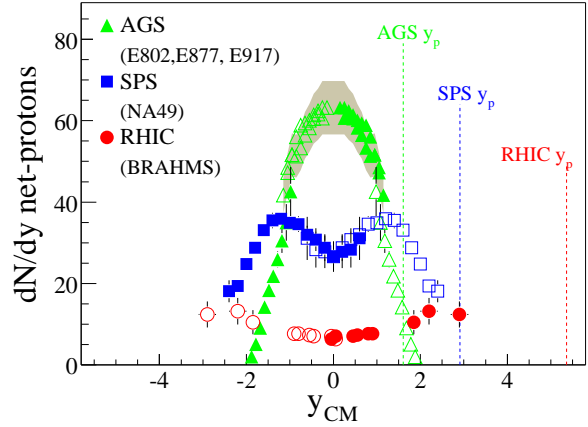}
  \caption{The net--proton rapidity distribution at
    AGS~\cite{e802,e877,e917} (Au+Au at \rootsnn{5}),
    SPS~\cite{Appelshauser:1998yb} (Pb+Pb at \rootsnn{17}) and this
    measurement (\rootsnn{200}). The data are all from the top 5\%
    most central collisions and the errors are both statistical and
    systematic (the light grey band shows the 10\% overall
    normalization uncertainty on the E802 points, but not the 15\% for
    E917). The data have been symmetrized. For RHIC data black points
    are measured and grey points are symmetrized, while the opposite
    is true for AGS and SPS data (for clarity). At AGS weak decay
    corrections are negligible and at SPS they have been applied.}
 \label{fig3}
\end{figure}

Figure~\ref{fig3} shows net--proton $dN/dy$ measured at AGS and SPS
compared to these results. The distributions show a strong energy
dependence, the net--protons peak at mid--rapidity at AGS, while at
SPS a dip is observed in the middle of the distribution. At RHIC a
broad minimum has developed spanning several units of rapidity,
indicating that at RHIC energies collisions are quite transparent.

To calculate the rapidity loss, $dN/dy$ must be known from
mid--rapidity to projectile rapidity, $y_{p} = 5.36$. BRAHMS measures
to $y \sim 3$, so the shape of the rapidity distribution must be
extrapolated to calculate $\drap$. The baryon number of participating
nucleons ($N_{part}$) is conserved, while the net--proton number is
not necessarily conserved. To obtain net--baryons, the number of
net--neutrons and net--hyperons have to be estimated and the
contribution from weak decays included in the measured net--protons
has to be deduced. Using MC simulations we find these contributions to
be $c_1 = 0.53 \pm 0.05$ protons for each $\Lambda$, and $c_2 = 0.49
\pm 0.05$ protons for each $\Sigma^+$ decay.  There is a weak rapidity
dependence which is included in the systematic error. The number of
net--baryons is, then,
\begin{equation}
  \label{eq:nbb}
  \nb = \npm \cdot \frac{\np + \nn + \nl + \nsp + \nsm}
       {\np + c_1 \nl + c_2 \nsp}, \nonumber
\end{equation}
\noindent
where $n_x = N_x - N_{\bar{x}}$ are the primary number of
net-neutrons ($\nn$), net-protons($\np$), net-lambdas($\nl$), and
net-sigmas($\nsm,\nsp$), respectively, and $\npm$ is the measured
net--proton yield. In addition to primaries, net-lambdas, $\nl$
includes contributions from other hyperons that decay to protons
through $\Lambda$s e.g. $\Sigma^0, \Xi^0, \text{and } \Xi^-$.

The ratio $\nnp = 1.00 \pm 0.05$ was found from
HIJING~\cite{Wang:1991ht} and AMPT~\cite{Zhang:1999bd} in the rapidity
interval $|y| < 3.5$.  The equilibration of protons and neutrons close
to mid--rapidity has been experimentally observed at AGS
energies~\cite{Barish:2001cf}.

At mid-rapidity the ratio $N_\Lambda/N_p = 0.89 \pm 0.07 \text{(stat)}
\pm 0.21 \text{(syst)}$ was found to be equal within statistical
errors to $N_{\bar{\Lambda}}/N_{\bar{p}} = 0.95 \pm 0.09 \text{(stat)}
\pm 0.22 \text{(syst)}$ at \rootsnn{130}~\cite{Adcox:2002au}. Those
results indicate that $N_\Lambda/N_p = N_{\bar{\Lambda}}/N_{\bar{p}} =
\nlp$. We use $\nlp = 0.93 \pm 0.11\text{(stat)} \pm 0.25
\text{(syst)}$ at \rootsnn{200} and find at $y=0$ the number of
feed-down corrected protons and anti--protons to be $dN/dy = 17.5 \pm
1.2 \text{(stat)} \pm 3.0 \text{(syst)}$ and $dN/dy = 13.2 \pm 0.9
\text{(stat)} \pm 2.3 \text{(syst)}$ respectively, in agreement with
the measurement at \rootsnn{200} by PHENIX : $dN/dy(p) = 18.4 \pm 2.6
\text{(syst)}, dN/dy(\bar{p}) = 13.5 \pm 1.8
\text{(syst)}$~\cite{Adler:2003cb}. Assuming that $\nlp$ is constant
over the full rapidity interval the feeddown correction for $\Lambda$
alone can be done as $\np = 0.67 \pm 0.12 \cdot \npm$. At forward
rapidity $\Lambda$s have not been measured, and the conservative
estimate $\nlp = 1.0 \pm 0.5$ is used.

The yield of $\Sigma^+$ and $\Sigma^-$ have not been measured at
RHIC. In thermal models $N_{\Sigma^+} \sim N_{\Sigma^-}$ and
$N_{\Sigma^-} \sim 0.1 \cdot N_p$~\cite{Broniowski:2001uk} at RHIC
(\rootsnn{130}), and $\nspp = \nsmp = 0.10 \pm 0.05$ is used here.

The final correction is $\nb = 2.03 \pm 0.08 \cdot \npm$. The
systematic error on the final correction depends almost entirely on
the error on $\nnp$ and $\nsmp$ because the partial derivative of
Eq.~\ref{eq:nbb} with respect to $\nlp$ and $\nspp$ is very small for
the numerical values used.

\begin{figure}[htb]
  \includegraphics[width=\columnwidth]{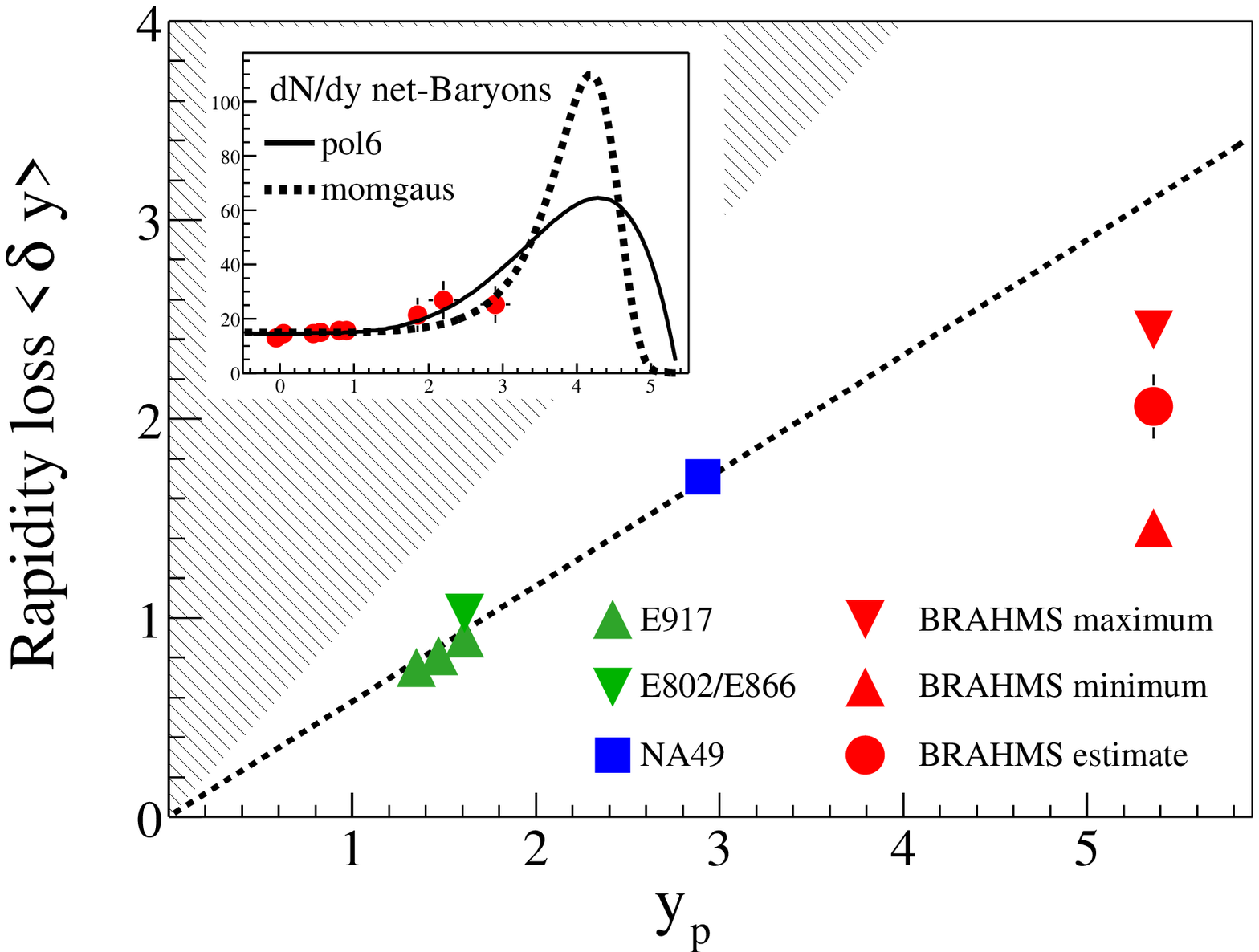}
  \caption{The inserted plot shows the extrapolated net--baryon
    distribution (data points) with fits (represented by the curves)
    to the data, see text for details. The full figure shows the
    rapidity loss, obtained using Eq.~\ref{eq:stop:baryonafter},
    as a function of projectile rapidity (in the CM).  The hatched
    area indicates the unphysical region and the dashed line shows the
    phenomenological scaling $\drap = 0.58 \cdot y_{p}$.  The data
    from lower energy are from~\cite{Videbaek:mf,e917}.}
 \label{fig4}
\end{figure}

Figure~\ref{fig4} (insert) shows the net--baryon $dN/dy$ obtained from
the measured net--proton $dN/dy$. The net--baryon density in the
rapidity region where BRAHMS has no acceptance can be constrained by
the total integral, equal to $N_{part}$, and since diffractive
scatterings can be ignored in central collisions where each nucleon on
the average experiences several scatterings, $dN/dy(y_p) \sim 0$.  To
illustrate the extrapolation from the measured region, $0 < y < 3$,
to beam rapidity, $y_p$, two different functions have been used: A six
order symmetric polynomial (pol6) $f(y)$ which is the simplest
polynomial that describes the data points, has the correct integral,
and $f(y_p)=0$, and a Bjorken inspired symmetric sum of 2 Gaussians in
momentum space (momgaus), converted to rapidity space using
$p=m \sinh(y)$ $(p_T=0)$ where $m$ is the proton mass. The momgaus
gives a simple description of the data as the product of the two
fragmenting nuclei with no mid-rapidity source. The data are
consistent with Bjorken's picture of the collision where the
fragmentation regions are far from mid--rapidity ($y \sim$ 3-4), but
shows that there is still a significant number of baryons transported
to mid--rapidity.  From the momgaus fit a mean momentum of 31 GeV/$c$
with a spread of 13 GeV/$c$ is found. The rapidity loss is $\drap =
2.04 \pm 0.10$ using the pol6 fit and $\drap = 2.06 \pm 0.16$ for the
momgaus fit, and the last value has been used as estimate in
Fig.~\ref{fig4}. From the pol6 fit there are $60 \pm 6$ net-baryons in
the rapidity interval, $0 < y < 3$, with a $\drap = 3.60 \pm 0.05$. If
the remaining 118.5 net-baryons are placed at y=3.5, maximum $\drap =
2.45$, and $y=5.0$, minimum $\drap = 1.45$, a conservative estimate
for the possible range of values is obtained, also shown in
Fig.~\ref{fig4}. The rapidity loss at RHIC is clearly less than the
phenomenological linear rapidity scaling would predict; this scaling
is broken at RHIC.

Using the functional forms the total energy, $E$, per net-baryon
after the collision can be derived:
 
\begin{equation}
  \label{eq:energy}
  E = \frac{1}{N_{part}}\int_{-y_{p}}^{y_{p}} \mmt \cdot
  \cosh{y} \cdot \frac{d\Nb}{dy} \cdot dy.
\end{equation}

\noindent
The $\mmt$ is known for protons from the spectra in the covered
rapidity interval.  Linear and Gaussian extrapolations of $\mmt$ to
projectile rapidity changes $E$ by less than 5\%. The energy is $E =
30 \pm 2$ GeV for pol6 and $E = 26 \pm 5$ GeV for the momgaus fits.
Taking $E = 27 \pm 6$, $\Delta E = 73 \pm 6$ GeV of the initial 100
GeV per participant is available for excitations. Using the same
minimum and maximum estimates above : $47 < \Delta E < 85$ GeV.

In conclusion BRAHMS has measured proton, anti--proton, and
net--proton yields from mid--rapidity ($y=0$) to forward rapidity ($y
\sim 3$). The net--proton distribution shows that collisions at RHIC
energies are quite transparent compared to lower energies. By
extrapolation to the full net--baryon distribution, we find that the
rapidity loss scaling observed at lower energy is broken and the
rapidity loss seems to saturate between SPS and RHIC energies.

This work was supported by the division of Nuclear Physics of the
Office of Science of the U.S. DOE, the Danish Natural Science Research
Council, the Research Council of Norway, the Polish State Committee
for Scientific Research and the Romanian Ministry of Research.

\newpage

\begin{thebibliography}{99}

\bibitem{Bass:1999zq}
  S.~A.~Bass {\it et al.},
  Nucl.\ Phys.\ A {\bf 661} (1999) 205.

\bibitem{Soff:2002bn}
  S.~Soff, J.~Randrup, H.~Stocker and N.~Xu,
  Phys.\ Lett.\ B {\bf 551} (2003) 115.
  
\bibitem{Bass:2002vm}
  S.~A.~Bass, B.~Muller and D.~K.~Srivastava,
  Phys.\ Rev.\ Lett.\  {\bf 91} (2003) 052302.
  
\bibitem{Busza:1983rj}
  W.~Busza and A.~S.~Goldhaber,
  Phys.\ Lett.\ B {\bf 139} (1984) 235.
  
\bibitem{Videbaek:mf}
  F.~Videbaek and O.~Hansen,
  Phys.\ Rev.\ C {\bf 52} (1995) 2684.
  
\bibitem{Hong:1997mr}
  B.~Hong {\it et al.}  [FOPI Collaboration],
  Phys.\ Rev.\ C {\bf 57} (1998) 244
  [Phys.\ Rev.\ C {\bf 58} (1998) 603].
  
\bibitem{e917}
  B.~B.~Back {\it et al.}  [E917 Collaboration],
  Phys.\ Rev.\ Lett.  {\bf 86} (2001) 1970.
  
\bibitem{Bjorken:1982qr}
  J.~D.~Bjorken,
  Phys.\ Rev.\ D {\bf 27} (1983) 140.

\bibitem{Karsch:2001vs}
  F.~Karsch,
  Nucl.\ Phys.\ A {\bf 698} (2002) 199.
  
\bibitem{BRAHMSNIM} 
  M.~Adamczyk {\it et al.} [BRAHMS Collaboration],
  Nucl.\ Instr.\ and Meth.\ A {\bf 499} (2003) 437.

\bibitem{Bearden:2001qq}
  I.~G.~Bearden {\it et al.}  [BRAHMS Collaboration],
  Phys.\ Rev.\ Lett.\  {\bf 88} (2002) 202301.
  
\bibitem{geant} GEANT 3.2.1, CERN program library.
  
\bibitem{Wang:1991ht}
  X.~N.~Wang and M.~Gyulassy,
  Phys.\ Rev.\ D {\bf 44} (1991) 3501.
  
\bibitem{Adler:2001aq}
  C.~Adler {\it et al.}  [STAR Collaboration],
Phys.\ Rev.\ Lett.\  {\bf 87} (2001) 262302


\bibitem{e802} L.~Ahle {\it et al.}  [E802 Collaboration],
  Phys.\ Rev.\ C {\bf 60} (1999) 064901.

\bibitem{e877}
  J.~Barette {\it et al.}  [E877 Collaboration],
  Phys.\ Rev.\ C {\bf 62} (2000) 024901.
  
\bibitem{Appelshauser:1998yb}
  H.~Appelshauser {\it et al.}  [NA49 Collaboration],
  Phys.\ Rev.\ Lett.\  {\bf 82} (1999) 2471.
  
\bibitem{Zhang:1999bd}
  B.~Zhang, C.~M.~Ko, B.~A.~Li and Z.~w.~Lin,
  Phys.\ Rev.\ C {\bf 61} (2000) 067901.

\bibitem{Barish:2001cf}
  K.~N.~Barish {\it et al.},
  Phys.\ Rev.\ C {\bf 65} (2002) 014904.

\bibitem{Adcox:2002au}
  K.~Adcox {\it et al.}  [PHENIX Collaboration],
  Phys.\ Rev.\ Lett.\  {\bf 89} (2002) 092302.

\bibitem{Adler:2003cb}
  S.~S.~Adler {\it et al.}  [PHENIX Collaboration],
  Submitted to Phys.\ Rev.\ C, arXiv:nucl-ex/0307022.
  
\bibitem{Broniowski:2001uk}
  W.~Broniowski and W.~Florkowski,
  Phys.\ Rev.\ C {\bf 65} (2002) 064905

\end{thebibliography}
\end{document}